\documentclass[twocolumn,prl,english]{revtex4-1}
\usepackage{pagenote}
\usepackage[T1]{fontenc}
\usepackage[latin9]{inputenc}
\usepackage{verbatim}
\usepackage{amsmath}
\usepackage{amssymb}
\usepackage{graphicx}
\usepackage{babel}
\usepackage[bottom]{footmisc}

\makeatletter

\usepackage{xcolor}
\newcommand*\diff{\mathop{}\!\mathrm{d}}

\begin{document}

\title{Motor-free contractility of active biopolymer networks}

\author{Sihan Chen$^{1,2}$, Tomer Markovich$^2$, Fred C. MacKintosh$^{1,2,3,4}$}
\affiliation{$^1$Department of Physics and Astronomy, Rice University, Houston, TX 77005\\
$^2$Center for Theoretical Biological Physics, Rice University, Houston, TX 77005\\
$^3$Department of Chemical and Biomolecular Engineering, Rice University, Houston, TX 77005\\
$^4$Department of Chemistry, Rice University, Houston, TX 77005}

\begin{abstract}
	Contractility in animal cells is often generated by molecular motors such as myosin, which require polar substrates for their function. Motivated by recent experimental evidence of motor-independent contractility,  we propose a robust motor-free mechanism that can generate contraction in biopolymer networks without the need for substrate polarity. We show that contractility is a natural consequence of {\it active} binding/unbinding of crosslinkers that breaks the \emph{Principle of Detailed Balance}, together with the asymmetric force-extension response of semiflexible biopolymers. 	
	We calculate the resulting contractile velocity using both a simple {coarse-grained} model and a more detailed microscopic model for a viscoelastic biopolymer network.  
	Our model may provide an explanation of recent reports of motor-independent contractility in cells. 
	Our results also suggest a mechanism for generating contractile forces in synthetic active materials. 
\end{abstract}
\maketitle
\section{Introduction}
Living cells are known to be far from equilibrium. Powered by metabolic components such as adenosine triphosphate (ATP), biophysical processes that cannot happen in equilibrium systems take place in living cells, including cell signaling~\cite{hernandez2012}, genetic transcription and replication~\cite{Murugan12034} and active force generation~\cite{Alberts}. Most force generation in living cells is due to molecular motors, for example myosin motors in animal cells~\cite{Fletcher2010485}. These motors perform directional motion on the substrate they bind to, thus generating force in various forms, from muscle contraction~\cite{Huxley1356} to internal stress of the cytoskeleton~\cite{Ingber2003, Marchetti2013,Markovich2019}. During cell division, motors are also believed to be responsible for the contractile stress generated by the actomyosin ring~\cite{Mabuchi1977,Guo1996,Straight2003,Glotzer2005}. 

Recent experimental evidence, however, suggests that contractility may be generated in the absence of motors~\cite{Wloka2013,Xue2016}. In Ref.~\cite{Wloka2013} it was found that myosin-II motors become immobilized shortly before cytokinesis in budding yeast. It was also shown that the contractility of the actomyosin ring in Drosophila embryo remains unaffected even when  motor activity is significantly inhibited~\cite{Xue2016}. Both experiments imply the existence of an underlying contractile mechanism other than molecular motors. 

\begin{figure*}[t]
	\centering
	\includegraphics[scale=0.6]{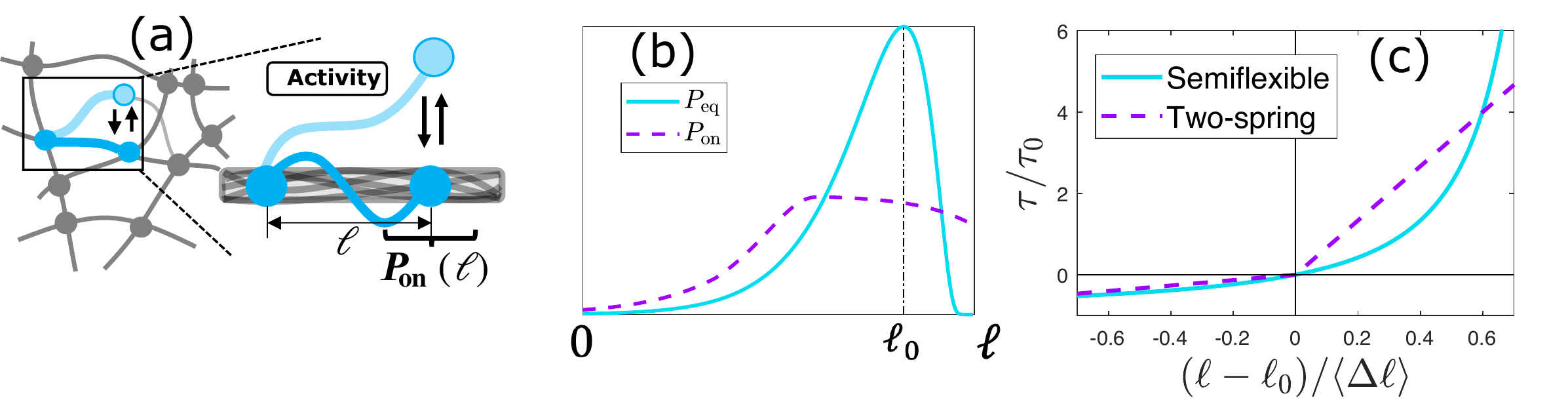}
	\caption{(a) Illustration of the coarse-grained model. One polymer segment from a disordered network undergoes binding/unbinding process attaining a steady-state length distribution, $P_{\rm on}$. 	
		The bound and unbound states are denoted by dark and light blue, respectively. 
		(b) Schematic plot (blue line) of the equilibrium distribution of a semiflexible polymer end-to-end length, $P_{\rm eq}$, and a sketch of a possible $P_{\rm on}$ (purple  dashed line). Activity usually broadens $P_{\rm on}$ as compared to $P_{\rm eq}$. 
		(c) Force-extension relation for inextensible semiflexible polymer ($\tau_0/\mu \to 0$ in Eq.~(\ref{e33})) and the two-spring PMF in Eq.~(\ref{e34}) with $K_1 = 2\tau_0/3\langle\Delta\ell\rangle$ and $K_2=10K_1$. Both PMFs show strong asymmetry, being potential sources of broken spatial symmetry. 
	}
	\label{fig_0}
\end{figure*}

Microscopically, molecular motors work with a mechanism reminiscent of the Smoluchowski-Feynman thermal ratchet~\cite{Feynman1963,Magnasco1993,Prost1994,Julicher1997}. By consuming and dissipating energy provided by metabolic components, motors undergo a directed cycle of transitions among conformational states~\cite{Spudich2001387}, while actively binding/unbinding in a way that violates the Principle of Detailed Balance (DB)~\cite{Boltzmann1872,Julicher1997, Battle2016604,Nardini2017,Markovich2020}. However, to generate directional motion or force, an energy consuming reaction itself is insufficient. It also requires a broken spatial symmetry which gives rise to a specific direction for the motion. This spatial asymmetry is usually due to the polarity of the substrate such as actin or microtubules, with well-defined plus and minus ends, to which a motor such as myosin or kinesin couples. Theoretically, such asymmetry  can be modeled as an asymmetric binding potential~\cite{Magnasco1993,Prost1994,Julicher1997}.

Another spatial asymmetry present in living cells that has been studied extensively, is the non-linear elasticity of the semiflexible biopolymers~\cite{Storm2005191,Gardel20061762,Mizuno2007}. The biopolymers that form the cytoskeleton are usually much stiffer to bending than most synthetic polymers, leading to a competition between the entropy and the bending energy. As a result, the force-extension relation of the biopolymers exhibits a strong asymmetry: when under stress, the polymer stiffness nonlinearly increases up to $10^3$ times~\cite{Bustamante19941599,Marko1994,Gardel20041301,Mizuno2007,Gardel20061762,Koenderink2009}, a phenomenon known as stress-stiffening.  However, under compression a $\rm pN$ level force is enough to buckle the polymer, and make its stiffness vanish~\cite{Broedersz2014}.  Such asymmetric force-extension relation, which originates from the thermal fluctuations of bending deformations, exists in most biopolymers including both polar filaments ({\it e.g.} actin and microtubule) and apolar filaments ({\it e.g.} intermediate filaments (IF)). This asymmetry in contraction vs expansion, is a potential source of broken spatial symmetry. 

In Ref.~{\cite{Chen2020}}, we have proposed a motor-independent contractile mechanism based on two key elements: (i) active binding/unbinding of non-motor crosslinkers which breaks DB and (ii) the asymmetric force-extension relation of biopolymers that prefers contraction over expansion. We have developed both a simple coarse-grained model and a detailed microscopic model to demonstrate the mechanism for a viscous or elastic substrate. In this paper, we extend our previous work~\cite{Chen2020} to include viscoelastic substrates and discuss the interplay between the binding/unbinding times and the substrate relaxation time.  Our proposed mechanism generates robust contractility for any viscoelastic substrates and any asymmetric force-extension. This model may not only provide a basis for understanding recent reports of myosin-independent contractility in cells, but also suggests a mechanism that can be used for active force generation in synthetic materials. 

\section{Overview}
In this work, we consider an unpolarized, disordered network formed by crosslinked semiflexible polymers (Fig.~\ref{fig_0}(a)) in which molecular motors cannot create contraction using their {\it power stroke}~\cite{Julicher1997}.  For simplicity, we focus on one particular polymer segment and treat the rest of the network as a continuum viscoelastic substrate that the particular polymer can bind (unbind) to (from). In the unbound state, there is no interaction between the polymer and the substrate, while in the bound state, the polymer exerts tension on the substrate, $\tau(\ell)= dU_e(\ell)/d\ell$, resulting in an average contractile force
\begin{equation}
\begin{aligned}
\langle F_s \rangle_\ell=\int P_{\rm on}(\ell)\tau(\ell)\diff\ell \, .
\end{aligned}
\label{e26}
\end{equation}
Here $\ell$ is the polymer end-to-end length, $U_e$ is the potential of mean-force (PMF) of the polymer, 
and $P_{\rm on}$ is the probability  for the polymer to have a given end-to-end distance $\ell$ in the bound state. If the binding and unbinding processes are in equilibrium, DB is satisfied and $P_{\rm on}(\ell)$ is governed by a Boltzmann distribution,  leading to vanishing contractile force. 
However, when the binding and/or unbinding process are out of equilibrium, {\it e.g.}, driven by consumption or catalysis of a metabolic component such as ATP,   
$P_{\rm on}$ can be non-Boltzmann. For a viscous substrate with drag coefficient $\gamma$,  the contractile force $\langle F_s\rangle_\ell$ creates an average contractile velocity, 
\begin{equation}
\begin{aligned}
v=\langle F_s \rangle_\ell/\gamma \, .
\end{aligned}
\label{e40}
\end{equation}

As discussed in the introduction, in order to have directed motion both time-reversal symmetry and spatial symmetry have to be broken~\cite{Feynman1963, Magnasco1993, Magnasco1994}. To see the manifestation of this principle within our model it is instructive to view the active process as an additional non-thermal noise on timescales longer than the binding/unbinding time~\cite{MacKintosh2008,Gladrow2016,Mura2018}, which causes random expansion or contraction of the polymer segment and breaks time-reversal symmetry. Such athermal noise usually tends to broaden the distribution $P_{\rm on}$~\footnote{\label{note1}we define the distribution width as the square root of the variance}, see  {\it e.g.}~Fig.~\ref{fig_0}(b). Notably, this active noise, which breaks time-reversal symmetry, cannot induce contraction or any directed motion without having some spatial symmetry broken~\cite{Feynman1963}. This can be seen from Eq.~(\ref{e26}), where symmetric $\tau$ together with symmetric $P_{\rm on}$   (from symmetry arguments all directions are the same) must lead to vanishing force. If $\tau$ (also the PMF, since $\tau$ is its derivative) is asymmetric, and specifically, soft to compression and hard to  extension, a symmetric or nearly symmetric $P_{\rm on}$ naturally leads to a positive contractile force. This can be understood intuitively: the polymer has equal chances to be extended or compressed by the active process. It exerts a large contractile force on the substrate when extended, while exerting a small expanding force when compressed, which on average leads to contractility. 

The PMF of semiflexible polymers has exactly these required properties: it stiffens under extension and softens under compression, see~Fig.\ref{fig_0}(c). The force-extension relation for a semi-flexible polymer is [Fig.~\ref{fig_0}(c)]~\cite{MacKintosh1995,Odijk1995,Wilhelm1996,Storm2005191,Broedersz2014}: 
\begin{equation}
\begin{aligned}
\frac{\ell(\tau)-\ell_0}{\langle\Delta\ell\rangle} = \frac{\ell_0 }{\langle\Delta\ell\rangle} \frac{\tau} {\mu} + \varepsilon \left( \frac{\tau}{\tau_0} \left(1+\frac{\tau}{\mu}\right)\right)
\, , 
\end{aligned}
\label{e33}
\end{equation}
where 
$\tau_0 \equiv \pi^2 k_B T  /  (6\langle\Delta\ell\rangle)$ is a characteristic tension, and $\langle\Delta\ell\rangle \simeq \ell_0^2/(6\ell_p)$ is the mean end-to-end thermal contraction of a stiff polymer of rest length $\ell_0$ and persistence length $\ell_p$.
Here, $\mu$ is the enthalpic stretch modulus, $k_B$ is the Boltzmann constant, $T$ is the temperature and $\varepsilon(\phi) = 1-3\frac{\pi \sqrt\phi \coth(\pi \sqrt\phi )-1} {\pi^2\phi}$~\cite{Broedersz2014}. As discussed above, the robustness of the contractility does not depend on the specific form of the asymmetric PMF. To illustrate this we will also use a simple and analytically tractable two-spring PMF
\begin{equation}
\begin{aligned}
\tau(\ell)=\left\{ 
\begin{aligned}
K_1(\ell-\ell_0) \qquad(\ell<\ell_0)\\
K_2(\ell-\ell_0) \qquad(\ell\geq \ell_0)
\end{aligned}
\right.
\end{aligned} \, ,
\label{e34}
\end{equation}
where $K_1<K_2$ are two different spring constants under compression and extension. 

The paper is organized as follows. We first detail our calculation of the contractile velocity for the coarse-grained model with a viscous substrate (Sec.~\ref{s4a}), which has been discussed in our previous work~\cite{Chen2020}. We then introduce a microscopic model which accounts for the details of the binding/unbinding process, and discuss the contractile velocity for a substrate characterized by the Maxwell model (Sec.~\ref{s3}). We show that, a viscoelastic substrate is equivalent to a viscous substrate with a modified binding potential (Sec.~\ref{s3}). Next, we discuss the conditions in which our microscopic model reduces to the coarse-grained one (see Sec.~\ref{s3}). In Sec.~\ref{s5} we draw our conclusions. 

\section{A Simple Coarse-grained Model}
\label{s4a}
We begin with a simple coarse-grained model to demonstrate how contractility can originate in an apolar substrate. As explained above, we consider a single polymer that binds to/unbinds from the substrate with rates $\omega_{\rm on}$/$2\omega_{\rm off}$ (the factor of $2$ accounts for the fact that both of the two polymer ends can unbind). When DB is obeyed the binding and unbinding rates are related by $\omega_{\rm on} =2\omega_{\rm off}\exp(-U_b)$, where $U_b$ is the binding potential which is usually rugged~\cite{Nishizaka1995,Julicher1997}. The rugged binding potential suggests that in DB $\omega_{\rm on}$ and/or $\omega_{\rm off}$ should vary according to the binding and unbinding position. Here we break DB in the simplest way by assuming that both $\omega_{\rm on}$ and $\omega_{\rm off}$ are constants~\cite{Julicher1997}. To start, consider a single binding/unbinding event that is divided into the following steps:
(i) in the unbound state the polymer length $\ell_u$ is assumed to relax fast to an equilibrium distribution with rest length $\ell_0$, $P_{\rm eq}(\ell_u)=\exp[-U_e(\ell_u)/k_B T]/Z$, where  $Z=\int \diff \ell_u\exp[-U_e(\ell_u)/k_B T]$ is the partition function  (Fig.~\ref{fig_0}(b)), 
(ii) the polymer binds to the substrate at rate $\omega_{\rm on}$; at the same time, its initial end-to-end length changes from $\ell_u$ to $\ell_b$ due to the binding, with probability $P_b(\ell_b)$, 
(iii) once the polymer binds, it contracts the viscous substrate due to its PMF, $U_e$, 
and (iv) the polymer {\it actively} unbinds at constant rate $2\omega_{\rm off}$. 
Here, we neglect thermal aspects of unbinding that we assume to be dominated by active processes. 

In this coarse-grained model, the details of the substrate binding potential are neglected. The effect of the binding process is modeled as an immediate change of the polymer length from $\ell_u$ (before binding) to $\ell_b$ (after binding). Such a change in length is a result of the rugged binding potential of biopolymers~\cite{Nishizaka1995,Julicher1997}. The probability that the polymer length has changed from $\ell_u$ to $\ell_b$ due to binding is denoted by $P_c(\ell_b,\ell_u)$, such that the binding probability (the polymer length distribution just after binding to the substrate) is:
\begin{align}
P_b(\ell_b) = \int \diff \ell_u P_{\rm eq}(\ell_u) P_c(\ell_b,\ell_u)\,.\label{e23}
\end{align}
In general the distribution $P_c$ depends on the form of both the binding potential and the elastic PMF. In the limit where the binding potential is much larger than PMF, $P_c$ is dominated by the substrate binding potential.  Then, for a homogeneous substrate $P_c$ is only a function of the length change (translational symmetry)~\cite{Chen2020}, $P_c(\ell_u,\ell_b)=P_c(\ell_u-\ell_b)$. Further assuming a symmetric binding potential, which is the case for {\it e.g.} isotropic substrate or a substrate consisting of apolar filaments, we find $P_c$  is symmetric around $\ell_u=\ell_b$. Note that we choose a symmetric binding potential as a `worst-case scenario' in which common motor activity is inhibited, although our mechanism does not rely on this symmetry. The simplest form of the binding probability is characterized by a single length scale $d$ that can be associated with the typical spacing between binding sites:
\begin{align}
P_c(|\ell_b-\ell_u|) = \frac 1 d \qquad{\rm for}\qquad  |\ell_b-\ell_u|< \frac d 2\,,\label{e36}
\end{align}
and $P_c=0$ otherwise.  In our model $P_c$ is a uniform distribution on  $(-d/2,d/2)$, but any symmetric (or even slightly asymmetric) $P_c$  leads to contractility. 

\begin{figure*}[t]
	\centering
	\includegraphics[scale=0.35]{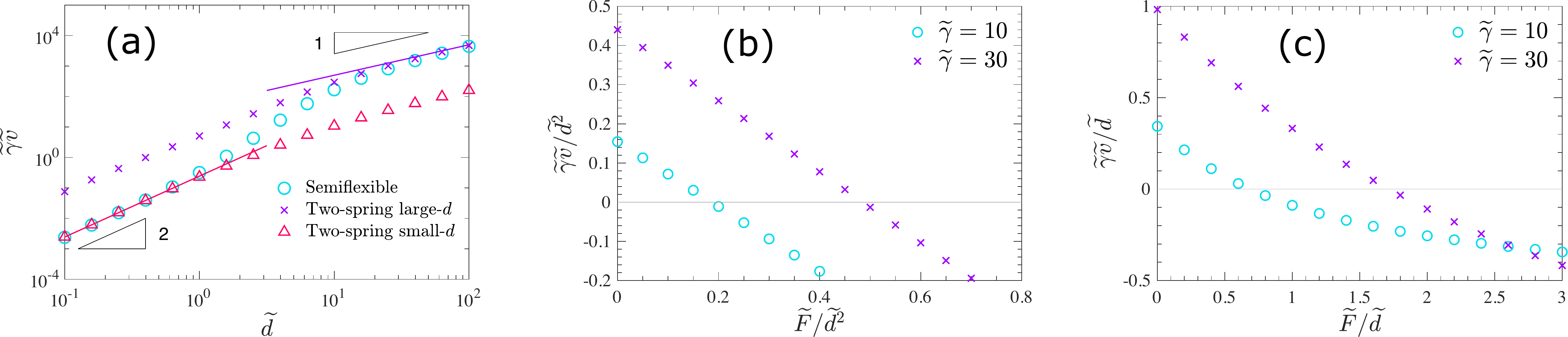}
	\caption{
		Numerical results for the contractile velocity.
		(a) Contractile velocity as function of the typical binding-site spacing for $\tilde \gamma\to \infty$. 
		For the semiflexible PMF, we use 
		$\mu = 4.37\times 10^{-8} \rm N$, $\tau_0 = 0.68 {\rm pN}$, $\ell_0 = 1\rm \mu m$, $\delta\ell= 6.2 {\rm nm}$, 
		while for the \emph{two-spring} PMF we choose $K_2 = \mu/\ell_0$ and $K_1 = 0.35\tau_0/\delta \ell$ (both PMFs are chosen to have the same $\delta \ell$). 		
		The solid purple line is the large-$d$ analytical solution of Eq.~(\ref{e29}) and the solid red line is the small-$d$ analytical solution of Eq.~(\ref{e38}). 
		In (b) and (c) we plot the force-velocity relation for the two-spring and the semiflexible PMFs, respectively.
		The dimensionless quantities used are: $\tilde d = d/\delta \ell$, $\tilde\gamma = 2\gamma\omega_{\rm off} \delta \ell /\tau_0$, 
		$\tilde v = v/ (2\omega_{\rm off}\delta \ell)$ and $\tilde F = F/\tau_0$. 
		Parameters used: (b) $\tilde d = 0.1$ , and (c) $\tilde d = 10$.  
		Both $\tilde F$ and $\tilde \gamma \tilde v$ are further rescaled according to their $d$ dependence (quadratic for large $\tilde d$ and linear for small $\tilde d$).
		In (a), (b) and (c) $\omega_{\rm on}\gg\omega_{\rm off}$. 	
	}
	\label{fig_min_v}
\end{figure*}

After the polymer binds to the substrate with initial length $\ell_b$ at $t=0$, it contracts the substrate due to its elastic energy.  The polymer length in the bound state, $\ell_f$, thus changes over time. For a viscous substrate with drag coefficient $\gamma$, the survival probability for $\ell_f$, $P(\ell_f;t)$ (probability that at time $t$ the polymer is still bound to the substrate and has length $\ell_f$), follows a 1-variable Fokker-Planck equation (FPE)
\begin{align}
&\partial_t {P}( \ell_f;t)+\partial_{\ell_f} {\bm {J}( \ell_f;t)}= - 2\omega_{\rm off}{P}\notag\\
& {P}( \ell_f;t=0)=P_b(\ell_b=\ell_f)\,,\label{e24}
\end{align}
where $J= -(k_B T/\gamma)\partial P/\partial \ell_f-{P}\partial U_e/\partial \ell_f $. The constant unbinding rate $2\omega_{\rm off}$ allows us to write $P$ as $P=P_s \exp(-2\omega_{\rm off})$, where the term $\exp(-2\omega_{\rm off})$ is the probability that the polymer is still bound at time $t$.  The distribution $P_s$ then satisfies a standard FPE:
\begin{align}
&\partial_t {P}_s( \ell_f;t)+\partial_{\ell_f} {\bm {J_s}( \ell_f;t)}= 0\notag\\
& {P_s}( \ell_f;t=0)=P_b(\ell_b=\ell_f)\,,\label{e35}
\end{align}
where $J_s= -(k_B T/\gamma)\partial{P_s}/\partial \ell_f-{P_s}\partial U_e/\partial \ell_f $.
Solving Eq.~(\ref{e35}) gives the distribution $P_s$, from which the steady state distribution $P_{\rm on}$ is calculated:
\begin{align}
&{P_{\rm on}}( \ell)=C_{\rm on}\int \diff t \cdot 2\omega_{\rm off} P_s(\ell_f=\ell;t)e^{-2\omega_{\rm off} t},\label{e25}
\end{align}
with $C_{\rm on} = \omega_{\rm on}/(\omega_{\rm on}+2\omega_{\rm off})$ being the probability to be in the bound state. We then find the contractile force, $\langle F_s\rangle_\ell$ through Eq.~(\ref{e26}), and the average contractile velocity through  Eq.~(\ref{e40}).

In Fig.~\ref{fig_min_v}(a) we numerically calculate the contractile velocity for both the semiflexible and the two-spring PMFs. We find positive contractile velocity for both PMFs, suggesting that our contractile mechanism is robust for any asymmetric elastic PMF that is hard to stretch and soft to compress. For both PMFs, the $d$-dependences of the velocity are similar. The velocity increases monotonically with $d$, with different scaling dependences in two regimes separated by a characteristic lengthscale $\delta \ell$. Here, $\delta \ell$ is defined to be the root-mean-square fluactuations of the polymer end-to-end  length under its equilibrium distribution. Its value is $\delta \ell=\sqrt{(\pi-2)k_B T/(\pi K_1)}$ for two-spring PMF (assuming $K_1\ll K_2$) and $\delta \ell =\ell_0^2/(\sqrt{90}\ell_p)$ for semiflexible PMF~(see supplementary material of Ref.~\cite{Chen2020}).  For $d\ll \delta \ell$, a quadratic $d$-dependence is observed, while for $d\gg\delta \ell$ we find that $v\sim d$. The different scaling exponents in these two limits originate in different profiles of $P_{\rm on}$: In the large-fluctuation limit ($\delta \ell\gg d$), $P_{\rm on}$ is only slightly perturbed from the equilibrium distribution $P_{\rm eq}$, while in the small-fluctuation limit ($ \delta \ell \ll d$), $P_{\rm on}$ is almost independent of $P_{\rm eq}$. Below we discuss in detail these two limits. 

\subsection{Small-fluctuation limit}
\label{s2a}
One can always write the corresponding Langevin equation of the FPE Eq.~(\ref{e35}) using standard methods~\cite{Gardiner}. For $\delta \ell \ll d$, thermal fluctuations are small enough that we can neglect the thermal term in the Langevin equation. In this case, for a given binding length $\ell_b$, the polymer length in the bound state is uniquely determined by a trajectory $\ell_f^*(\ell_b;t)$ which follows a  Langevin equation that corresponds to Eq.~(\ref{e35}):
\begin{align}
\gamma \frac{\diff{\ell}_f^*(\ell_b;t)}{\diff t} &= -\tau\left(\ell_f^*\right)\notag\\
{\ell}_f^*(\ell_b;t=0)&=\ell_b\, ,\label{e27}
\end{align}
where $\tau (\ell_f)=U_e'(\ell_f)$ is the polymer tension.  In Eq.~(\ref{e35}), for a given binding length $\ell_b$, {\it i.e.,} an initial condition $P_s(\ell_f;t=0 ) = \delta (\ell_f-\ell_b)$, the trajectory in Eq.~(\ref{e27}) gives a solution of Eq.~(\ref{e35}), $P_s = \delta(\ell_f-\ell_f^*)$, where $\delta(x)$ is the Dirac $\delta$-function. Therefore, the resulting $P_s$ from the binding probability $P_b(\ell_b)$, is the average probability distribution of trajectories starting with all possible $\ell_b$
\begin{align}
P_s(\ell_f;t) =\int \diff \ell_bP_b(\ell_b) \delta\left[\ell_f-\ell_f^*(\ell_b;t)\right]\, .\label{e28}
\end{align}

The steady-state distribution in the bound state is then calculated by solving Eq.~(\ref{e28}) and substituting  $P_s$ into Eq.~(\ref{e25}). Then, the contractile velocity can be found with the help of Eq.~(\ref{e26}) and Eq.~(\ref{e40}).  For the two-spring PMF, this velocity assumes a simple form:
\begin{align}
v=&\frac{dC_{\rm on}\gamma \omega_{\rm off}^2(K_2-K_1)}{8(K_1+\gamma \omega_{\rm off})(K_2+\gamma \omega_{\rm off})} \,,\label{e29}
\end{align}
which depends linearly on $d$. This expression is in perfect agreement with the numerical solution of the FPE, Eq.~(\ref{e35}), see Fig.~\ref{fig_min_v}(a). 

The contractile velocity for the semiflexible potential cannot be found analytically, because the potential does not have an explicit analytical expression. However, in the $\delta \ell \ll d$ limit, the force-extension relation of the semiflexible PMF has two asymptotic limits, for large stretch or compression~\cite{Broedersz2014}: 
\begin{equation}
\begin{aligned}
\tau(\ell)\approx
\left\{ 
\begin{aligned}
&\frac{\mu}{\ell_0}(\ell-\ell_0) \qquad (\ell-\ell_0\gg\delta\ell)\\
&-\tau_0 \qquad \qquad \,\,(\ell_0-\ell\gg\delta\ell) \, .
\end{aligned}
\right.
\end{aligned}
\label{e30}
\end{equation}
Therefore, if we approximate the semiflexible PMF by a two-spring PMF with $K_2 = \mu/\ell_0$ and $K_1\ll K_2$, the two PMFs generate the same average force  just after binding ($K_1$ does not affect the average force as long as $K_1\ll K_2$). In Fig.~\ref{fig_min_v}(a) we show that the contractile velocity of the semiflexible PMF coincides with that of the two-spring PMF in the small-fluctuation (large-$d$) limit, given the proper choice of the two-spring parameters. To be specific, we let the two PMFs have the same $\delta \ell$. For this aim we set $K_1 = 3.32\tau_0\ell_p/\ell_0^2=0.35\tau_0/\delta \ell$ such that the two spring PMF gives the same $\delta \ell$ with the semiflexible PMF. 

\subsection{Large-fluctuation limit} 
For $\delta \ell \gg d$, thermal noise cannot be neglected, but the change in length due to the binding is relatively small. Then, $P_b$ is only slightly perturbed from the equilibrium distribution in the unbound state. Expanding $P_b$ of Eq.~(\ref{e23}) to second order in $d$ gives
\begin{align}
P_b(\ell_b)\simeq
P_{\rm eq}(\ell_b)\left[1+\frac{d^2}{24}\left(\frac{U_e'^2(\ell_b)}{(k_BT)^2}-\frac{U_e''(\ell_b)}{k_B T}\right)\right]\,,\label{e31}
\end{align}
where $f'(\ell_b)=[\diff f(\ell) /\diff \ell]_{\ell=\ell_b}$ and $f''(\ell_b)=[\diff^2 f(\ell) /\diff \ell^2]_{\ell=\ell_b}$. Since both the FPEs of Eq.~(\ref{e35}) and Eq.~(\ref{e25}) are linear, the resulting $P_{\rm on}$ is similarly perturbed around the equilibrium distribution, where the deviation scales as $d^2$. Thus, from  Eq.~(\ref{e26}) we have $\langle F_s \rangle_\ell \sim d^2$, and the contractile velocity shows quadratic dependences on $d$, as is seen in Fig.~\ref{fig_min_v}(a). To show this explicitly, let us consider a nearly rigid substrate, {\it i.e.}, $ \tilde \gamma \rightarrow \infty$. 
In this case the polymer is not relaxing in the bound state ({\it i.e.,} $P_{\rm on}(\ell) = C_{\rm on} P_b(\ell_b=\ell)$), and produces an average contractile force (on the substrate) of
\begin{align}
\langle F_s \rangle_\ell =\frac{C_{\rm on}d^2}{24}\int \diff \ell_b P_{\rm eq}(\ell_b)U_e'''(\ell_b) \, ,
\label{e37}
\end{align}
where we have used Eqs.~(\ref{e26}) and (\ref{e31}). This force is positive for any potential with positive $U_e'''(\ell)$. 
Note that the above result is obtained for constant on/off rates. 
In case that the on/off rates obey detailed balance,  $P_{\rm on}$ would remain the equilibrium distribution even for finite $d$ and $\langle F_s \rangle_\ell$ would vanish. For the two-spring PMF, the contractile force calculated from Eq.~(\ref{e37}) leads to a contractile velocity:
\begin{align}
v =\frac{C_{\rm on}(K_2-K_1)\sqrt{K_1}d^2}{12\sqrt{2\pi k_B T}\gamma} \, ,
\label{e38}
\end{align}
which agrees perfectly to the numerical results in the small-$d$ limit, see Fig.~\ref{fig_min_v}(a). For the semiflexible PMF, numerical result gives $v\approx0.24d^2\tau_0/(\delta\ell^2\gamma)$. Comparing this result with Eq.~(\ref{e38}), we find that by setting $K_1 = 0.35\tau_0/\delta\ell$ and $K_2 = 13\tau_0/\delta\ell$, the two-spring PMF gives the same contractile velocity with the semiflexible PMF in the small $d$ limit, which is confirmed by our numerical results, see Fig.~\ref{fig_min_v}(a). Here, the value of $K_1$ is chosen to be the same as that in Sec.~\ref{s2a}, such that the two-spring PMF has the same $\delta \ell$ with the semiflexible PMF. We then set the value of $K_2$ such that Eq.~(\ref{e38}) gives the same velocity as the semiflexible PMF. Together with previous discussion on the small-fluctuation limit, this suggests that given the proper parameter choice, the two-spring PMF can mimic the semiflexible PMF both in the small and large-fluctuation limits.

Having illustrated the coarse-grained model in two extreme limits, let us consider a property that is commonly measured for molecular motors, the force-velocity relation. This relation describes a motor's ability to do work under an external load. Since our non-motor mechanism creates motor-like contraction, it is also useful to compute the force-velocity relation in our model. To calculate this relation, we exert a constant tension $F$ on the two ends of the polymer. The constant force modifies the elastic potential of Eqs.~(\ref{e24}) and (\ref{e35})  from $U_e(\ell_f)$ to $U_e(\ell_f)-F\ell_f$. In Fig.~\ref{fig_min_v}(b)-(c) we plot force-velocity curves for the small and large $d$ limits, respectively. 
In Fig.~\ref{fig_min_v}(b) we use the two-spring PMF, while in Fig.~\ref{fig_min_v}(c) we use the semiflexible polymer PMF. 
In both cases, the velocity is reduced by an applied load, in a way similar to molecular motors~\cite{Derenyi19966775}. 
We find that decreasing the dimensionless viscosity $\tilde \gamma$ results in lower velocities and correspondingly lower stall forces,
 due to the increased compliance and stress relaxation of the substrate. 
 
\section{Microscopic Model}
\label{s3}
\begin{figure*}[t]
	\centering
	\includegraphics[scale=0.33]{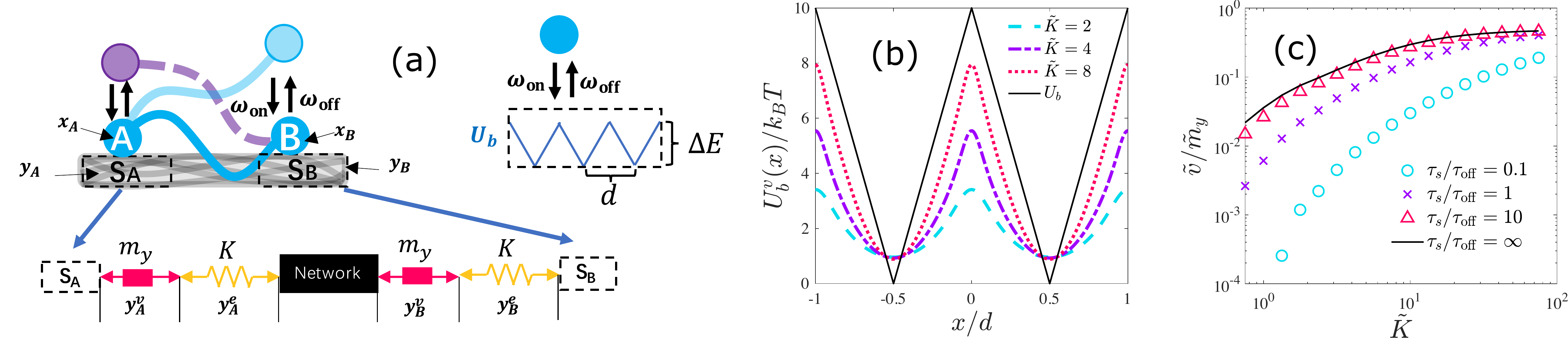}
	\caption{(a) Illustration for the microscopic model. Two ends of a polymer actively bind to and unbind from two regions of a Maxwell substrate. Each of the substrate regions is connected to the rest of the network via a viscous damper and an elastic spring.  Each of the polymer ends binds to the substrate region with a triangular binding potential (upper right). (b) Profile of the modified binding potential $U_b^v$  for various rescaled substrate spring constants, $\tilde K = K d^2/\Delta E$. The height of $U_b^v$ increases with $\tilde K$ and reaches $U_b$ when $\tilde K \to \infty$. (c) Rescaled contractile velocity, $\tilde v = v /(\omega_{\rm off }\delta \ell)$, as function of $\tilde K$, for various ratios between the substrate relaxation time $\tau_s=1/(Km_y)$ and the unbinding time $\tau_{\rm off}$. The velocity increases monotonically with $\tilde K$ and approaches its maximum as $\tilde K \gg 1$. The PMF is the same as the semiflexible PMF used in Fig.~\ref{fig_min_v}.  In (c) $d=10 \rm nm$.   }
	\label{fig_ve}
\end{figure*}
Having demonstrated our mechanism using a simple coarse-grained model, let us switch to a more complete microscopic model. We will later show how the coarse-grained model can be derived as a limit of this microscopic model.  In this model, we still study the active binding/unbinding of a semiflexible segment on a viscoelastic substrate where the details of the binding potential will be considered.  

As sketched in Fig.~\ref{fig_ve}(a), we consider a semiflexible segment whose two ends, A and B, bind and unbind to two regions on a viscoelastic substrate denoted as $\rm S_A$ and $\rm S_B$. Each of the substrate regions represents a small part of the entire substrate, and they can move independently to each other or the rest of the substrate. The substrate regions are assumed to be rigid and their positions are denoted by $y_A$ and $y_B$, respectively. We consider a Maxwell substrate by connecting each substrate region with the rest of the substrate with a viscous damper of mobility $m_y$ and a spring of spring constant $K$ in series. Let $y_{A}^e$ and $y_{A}^v$ be the lengths of the elastic and viscous parts of the substrate region $\rm S_A$, and similarly define $y_{B}^e$ and $y_{B}^v$ (note that these lengths can be negative as they are measured relative to the initial positions). These lengths satisfy $y_{A,B} = y_{A,B}^e + y_{A,B}^v$. 

The two polymer ends (crosslinkers) can be thought as particles moving within the substrate with effective mobility $m_x$, whose positions are denoted by $x_A$ and $x_B$, respectively. The polymer end A (B), successively binds to and unbinds from the substrate region $\rm S_A$ ($\rm S_B$). When A (B) is unbound, there is no interaction between the polymer end A (B) and the substrate region $\rm S_A$ ($\rm S_B$).  When A (B)  is bound to $\rm S_A$ ($\rm S_B$), the polymer end interacts with the substrate region via an effective binding potential, $U_b(x_A-y_A)$ ($U_b(x_B-y_B)$). 
The binding potentials of biopolymers are usually rugged due to the microscopic structure of the monomers. Since we assume the substrate is apolar, these potentials should  be symmetric. For simplicity, we assume $U_b$ to be a periodic triangular potential of depth $\Delta E$ and period $d$, although periodicity is not essential. Lastly, the PMF of the polymer is $U_e(x_B-x_A)$. 

The binding and unbinding rates of the two polymer ends are denoted by $\omega_{\rm on}$ and $\omega_{\rm off}$, respectively. As discussed in the beginning of Sec.~\ref{s4a}, constant binding/unbinding rates break DB, which together with a rugged binding potential and asymmetric elasticity can lead to a steady state substrate contraction.

The polymer can only exert tension on the substrate when its two ends are bound to the substrate, and we define this state as the bound state. The unbound state is thus the state in which one or both polymer ends are unbound. We are interested in how the substrate length, defined as $y_B-y_A$, changes during the bound state. For that aim we consider a bound state which starts at $t=0$. Just before $t=0$, one of the two ends must be bound to the substrate region while the other being unbound, so the system could enter the bound state. We assume the polymer end A is bound while B is unbound (this is equivalent to the case in which B is bound while A is unbound). At $t=0$, B binds to $S_B$, and the system enters the bound state. While in the bound state, the polymer interacts with the substrate and deforms it. The evolution of the system is described by a survival probability ${\cal P}^{\rm ve}\left(x_{A,B}, y_{A,B}^e,y_{A,B}^v;t\right)$ (each variable with subscript $A,B$ stands for two variables with substripts $A$ and $B$, respectively).  The bound state ends at $t=t_e$, when one of the polymer ends unbinds from the substrate. The distribution of $t_e$ is exponential with unbinding rate $2\omega_{\rm off}$ (derivation is as in the paragraph before Eq.~(\ref{e35})), where the factor of two is due to the fact that the unbinding of either polymer ends terminates the bound state. 

During a single binding/unbinding event, the average contraction of the substrate is defined as
\begin{align}
\Delta y  =  \langle y_B^v-y_A^v\rangle _{t=0}-\langle y_B^v-y_A^v\rangle _{t=t_e}\,,
\label{e1}
\end{align} 
where
\begin{align}
\langle y_B^v-y_A^v\rangle _{t=0}&=\int DxDy\, \, (y_B^v-y_A^v)\notag\\&\times{\cal P}^{\rm ve}\left(x_{A,B}, y_{A,B}^e,y_{A,B}^v;t=0\right)\,\notag\\
\langle y_B^v-y_A^v\rangle _{t=t_e}&=2\omega_{\rm off}\int_0^\infty \diff t_e\int DxDy\, (y_B^v-y_A^v)\, \, \notag\\&\times{\cal P}^{\rm ve}\left(x_{A,B}, y_{A,B}^e,y_{A,B}^v;t=t_e\right)\,\,\label{e39}
\end{align} 
are the average values of $y_B^v-y_A^v$ at the start and at the end of the bound state. 
Here $Dx=\diff x_A\diff x_B $ and $D_y=\diff y_A^e\diff y_B^e \diff y_A^v \diff y_B^v$. In Eq.~(\ref{e39}) we only consider the viscous parts of the substrate, because the elastic parts relax immediately to their rest lengths in the unbound state, and the lengths of the viscous parts are the only lengths that show changes due to the relaxation in the bound state. 

The contractile velocity is calculated using $v =  \Delta y/ {\cal T}$, where ${\cal T}$ is the average time between two binding/unbinding events. By definition, ${\cal T}$ is the sum of the average lifetimes in the bound and unbound states. The average lifetime of the polymer in the bound state (both ends are bound) is $\tau_{\rm off}=1/2\omega_{\rm off}$. The average lifetime in the unbound state is more complicated as it is composed of two different states: (a) both ends are unbound, and (b) only one end is unbound. The probabilities to be  in these two states can be written as, $C_a = (1-C_{\rm on})^2$, $C_b = 2C_{\rm on}(1-C_{\rm on})$, respectively. Here $C_{\rm on}=\omega_{\rm on}/(\omega_{\rm on}+\omega_{\rm off})$ is the fraction of time in which one of the polymer ends is bound to the substrate (regardless of the other polymer end), see also Eq.~(\ref{e25}). Therefore, when the polymer is in the bound state, the probabilities to be in states (1) and (2) are, $P_a=C_a/(C_a+C_b)$ and $P_b=C_b/(C_a+C_b)$, respectively. Since the unbound state can only be ended when the system is in state (2), the net binding rate, given that the polymer is in the unbound state, is $\omega^*_{\rm on}=P_b\omega_{\rm on}$, and the average lifetime of the unbound state is $\tau_{\rm on} = 1/\omega^*_{\rm off}$. Taken together, we have ${\cal T} = \tau_{\rm off}+\tau_{\rm on}=1/(2C_{\rm on}^2 \omega_{\rm off})$.

In the bound state, the evolution of the system can be described by six variables: $x_{A,B}$, $y_{A,B}^e$ and $y_{A,B}^v$, with the total potential energy $W^{\rm ve}\left(x_{A,B},y_{A,B}^e,y_{A,B}^v\right)$: 
\begin{align}
W^{\rm ve} &=  U_e(x_B-x_A)+U_b(x_B -y_B^e-y_B^v)\notag\\&+U_b(x_A-y_A^e-y_A^v)+\frac{K}2 (y_A^e)^2+\frac{K}2 (y_B^e)^2\,.\label{e9}
\end{align}
Note that we set the rest lengths of both $y_A^e$ and $y_B^e$ to zero for simplicity.

In general, the survival probability of the system should also be described by these 6 variables, ${\cal P}^{\rm ve}\left(x_{A,B},y_{A,B}^e,y_{A,B}^v;t\right)$. However, we can eliminate $y_A^e$ and $y_B^e$ since they obtain their deformed values instantaneously, such that they can always be considered as fast variables. Therefore, we can describe the system dynamics using a 4-variable survival probability ${\cal P}^v$~\cite{Magnasco1994}:
\begin{align}
{\cal P}^{\rm ve} &=  \frac{\exp[-W^{\rm ve}/k_B T]}{Z^v\left(x_{A,B}, y_{A,B}^v\right)} {\cal P}^{v}\left(x_{A,B}, y_{A,B}^v;t\right)\,.\label{e10}
\end{align}
with $Z^v=\int \diff y_A^e \diff  y_B^e \, {\exp({- W^{\rm ve}/k_BT})}$. 

 The evolution of ${\cal P}^v$ satisfies a FPE, 
\begin{align}
\partial_t {\cal P}^v+\nabla \cdot { {\bm J^v}}= - 2\omega_{\rm off} {\cal P}^v\,,\label{e11}
\end{align}
with $J_\alpha=-m_\alpha( k_B T\partial_\alpha {\cal P}^v+{\cal P}^v  \partial_\alpha W^v)$ without the summation convention. Here $m_\alpha=m_x$ for $\alpha=x_{A,B}$ and  $m_\alpha=m_y$ for $\alpha=y_{A,B}^v$. Notably, this FPE is governed by an effective potential $W^v\left(x_{A,B}, y_{A,B}^v\right)$: 
\begin{align}
W^v&=\int \diff y_A^e \diff y_B^e \frac{  \exp({- W^{\rm ve}/k_BT})}{Z^v\left(x_{A,B}, y_{A,B}^v\right)}\notag\\&\times W^{\rm ve}\left(x_{A,B}, y_{A,B}^e,y_{A,B}^v\right)\notag\\&\!\!\!\!\!\!=U_e(x_B-x_A)+U_b^v(x_B-y_B^v)+U_b^v(x_A-y_A^v)\,.\label{e12}
\end{align}
Surprisingly, we find this effective potential to have the same form as the original total potential energy of a pure viscous substrate (see discussion later on the viscous substrate), but with a modified binding potential
\begin{align}
U_b^v(x)&=\int \diff y\frac{\exp\left[-[U_b(x-y)+\frac K 2 y^2]/k_B T\right]}{Z_b(x)}\notag\\&\times \left[U_b(x-y)+\frac K 2 y^2\right]\,,\label{e13}
\end{align}
where $Z_b(x)=\int \diff y\,{\exp\left[-[U_b(x-y)+ K  y^2/2]/k_B T\right]}$. This suggests that a viscoelastic substrate is equivalent to a viscous substrate with a modified binding potential $U_b^v$. 

The effective potential $U_b^v$ couples the original binding potential $U_b$ and the elastic energy of the substrate. Same as the bare binding potential, it is periodic with period $d$, but the profile of each potential well is flattened by the elastic energy (see Fig.~\ref{fig_ve}(b)). For $K\gg \Delta E/d^2$, the substrate shows almost a pure viscous response, where $U_b^v\approx U_b$. For smaller $K$, $U_b^v$ deviates from $U_b$, and for $K\ll \Delta E/d^2$, $U_b^v$ becomes entirely flat. This is because when $K\to\infty$ the Maxwell substrate reduces to a pure viscous substrate, and  when $K$ is small the elastic compliance of the substrate is weak and almost all the stress can be relaxed by the deformation of the elastic part, while the deformation of the viscous part is negligible.   Therefore, in order to generate non-trivial contractility during the binding/unbinding events, the substrate rigidity $K$ should be large enough ($K\geq \Delta E/d^2$).

To determine the initial condition of Eq.~(\ref{e11}), we need to specify the meaning of $y_A^v$ and $y_B^v$. Since the substrate regions are defined to be rigid, their positions can be represented by any point within them. Here, for simplicity we choose $y_A^v$ and $y_B^v$ to be the positions of the nearest binding sites (bottoms of the potential wells in $U_b^v$) of A and B at $t=0$, respectively. Hence, at $t=0$ we have $|x_{A,B}-y_{A,B}^v|\leq d/2$. Assuming the system relaxes fast in the unbound state, the initial condition is given by
\begin{align}
&{\cal P}^v\left(x_{A,B}, y_{A,B}^v;t=0\right)=\frac{\chi\left(x_{A,B}, y_{A,B}^v\right)}{ Z}\notag\\&\times\exp{\Big(- [U_e(x_B-x_A)+U_b^v(x_A-y_A)]/k_B T\Big)}\,,\label{e4}
\end{align}
where $Z = \int DxDy^v {\cal P}^v$ is the partition function ($Dx = \diff x_A \diff x_B$ and $Dy^v = \diff y_A^v \diff y_B^v$) and 
\begin{align}
\chi\left(x_{A,B}, y_{A,B}^v\right)&= \Theta\left(d/2 - \left|x_A - y_A^v\right|\right)\notag\\&\times\Theta\left(d/2 - \left|x_B - y_B^v\right|\right)\,,\label{e5}
\end{align}
gives the  boundaries of the initial condition ($\Theta(x)$ is the Heaviside function). $\chi$ appears because we define the values of $y_A^v$ and $y_B^v$ at $t=0$ to be the positions of the nearest binding sites of $\rm A$ and $\rm B$. Note that $U_b^v$ for the polymer end B is taken into account in the initial condition, because it is assumed to be bound and immobile during the binding of A. 

The survival probability at time $t$  is then derived from Eq.~(\ref{e11}) and (\ref{e4}), which can  be used to calculate the contractile velocity. The large number of variables in the FPE prevents intuitive understanding of the dynamics, and further introduces difficulties in the numerical solution. Below we will apply two physical conditions that reduces the 4-variable FPE  to a 1-variable equation. 

First, we assume $m_x \gg m_y$ because $m_x$ is the mobility of one end of a single polymer, while $m_y$ stands for the mobility of a substrate region which is a collection of multiple polymers that feels larger friction than a single polymer end. 

To introduce the second condition we consider three key timescales in the system. The first one is the time for a single polymer end to relax within one potential well, $\tau_{r} = \tau_d\left({k_B T}/{\Delta E^v}\right)$, where $\tau_d = { d^2}/{2k_B Tm_x}$ and $\Delta E^v$ is the difference between the maximum and minimum of $U_b^v$. The complete derivation of $\tau_r$, which is the so-called intrawell relaxation time~\cite{Derenyi1999} is provided in the Appendix. The second timescale is the time for the polymer end to hop to another potential well, $\tau_{\rm hop} =  \tau_d \exp\left({\Delta E^v}/{k_B T}\right)$ (see Appendix for derivation). Lastly, the third timescale is the average lifetime in the bound state, $\tau_{\rm off} = 1/(2\omega_{\rm off})$. In biopolymers $\Delta E$ is usually much larger than $k_B T$, and we assume $\Delta E^v$ is also large enough such that $\tau_r \ll \tau_{\rm off} \ll \tau_{\rm hop}$. This requires the network rigidity $K$ to be comparable or larger than $\Delta E/d^2$. Because the value of $K$ is in principle proportional to the network density, the assumption is relevant for densely-crosslinked network. The relation between the three timescales implies that during the bound state the polymer end relaxes quickly within the initial potential well and unbinds before it can hop to another potential well. 

\begin{figure*}[t]
	\centering
	\includegraphics[scale=0.32]{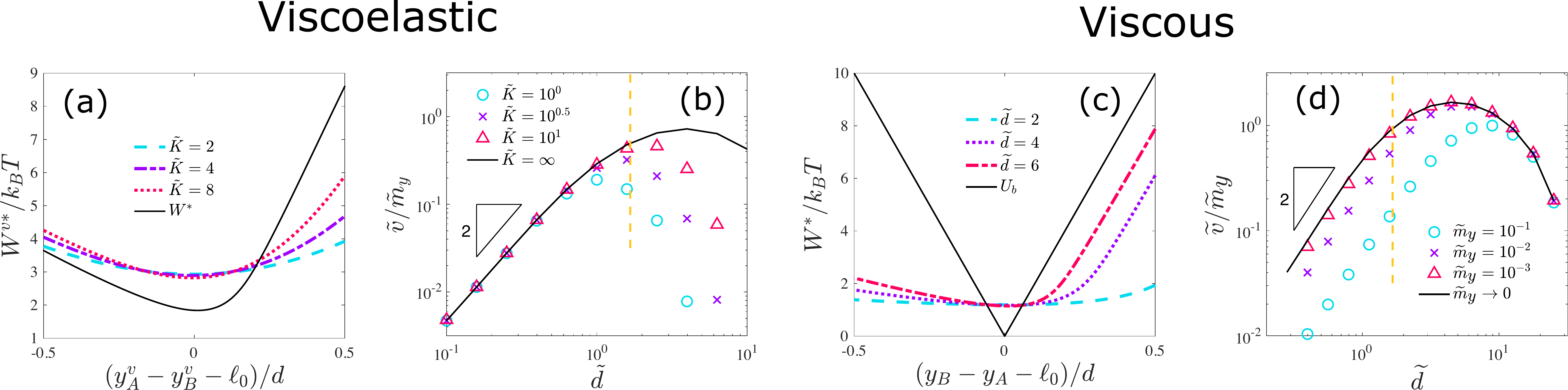}
	\caption{Numerical results for the microscopic model. (a) Profile of the modified effective interaction $W^{v*}$, for various rescaled substrate spring constants, $\tilde K = K d^2/\Delta E$. $W^{v*}$ becomes more asymmetric for increasing $\tilde K$ and approaches $W^*$ of the viscous substrate limit (see (c)) when $\tilde K \to \infty$. (b)  Rescaled contractile velocity, $\tilde v = v /(2\omega_{\rm off }\delta \ell)$, as function of rescaled binding site spacing $\tilde d$, for various $\tilde K$ values. The dependence on $\tilde d$ is non-monotonic. In the small $\tilde d$ limit the microscopic model reduces to the coarse-grained model. The maximum velocity is reached at $\tilde d \sim 1$. Dashed yellow line shows a realistic value $d=10 \rm nm$ with $\delta \ell = 6\rm nm$, which is close to the maximum-velocity location.  (c) Profile of the effective interaction $W^{*}$ in the viscous limit, for various $\tilde d$ values. $W^{*}$ is more asymmetric for small $d$ and approaches $U_b$ when $\tilde d \to \infty$. (d)  Rescaled contractile velocity as function of rescaled binding site spacing  for various rescaled substrate mobility, $\tilde m_y = m_y\tau_0/(2\omega_{\rm off}\delta \ell)$ values in the viscous substrate limit. Similarly to the viscoelastic case (see (b)), the dependence on $d$ is non-monotonic. PMF parameters are the same as that in Fig.~\ref{fig_min_v}. The binding potential is the same as that in Fig.~\ref{fig_ve}.      }
	\label{fig4}
\end{figure*}

  With these two physical conditions we simplify the FPE as follows. Since $\tau_{\rm off} \ll \tau_{\rm hop}$, the probability that either of the polymer ends hops before unbinding is small, and both polymer ends are trapped in the potential wells to which they bind,  indicating that ${\cal P}^v\sim \chi\left(x_{A,B}, y_{A,B}^v\right)$. Furthermore, because $\tau_r \ll \tau_{\rm off}$  and $m_x \ll m_y$, we can treat $x_A$ and $x_B$ as fast variables, such that
\begin{align}
\!{\cal P}^v\left(x_{A,B}, y_{A,B}^v;t\right)&\!\!=\!\frac{\exp{\left[- W^v\left(x_{A,B}, y_{A,B}^v\right)\!\!/k_B T\right]}}{Z_Y\left(y_{A,B}^v\right)}\notag\\&\!\!\!\!\!\!\!\!\!\!\!\!\!\!\!\!\!\!\!\!\!\!\!\!\!\!\!\!\!\!\times\chi\left(x_{A,B}, y_{A,B}^v\right){\cal P}^v_Y\left(y_{A,B}^v;t\right)\,,\label{e6}
\end{align}
with $Z_Y=\int \diff x_A \diff  x_B \chi \, {e^{- W^v/k_BT}}$. The distribution ${\cal P}^v\left(x_{A,B}, y_{A,B}^v;t\right)$ can thus be determined by its marginal distribution ${\cal P}^v_Y\left(y_{A,B}^v;t\right)$. Substituting Eq.~(\ref{e6}) into Eq.~(\ref{e11}), we have the reduced FPE for ${\cal P}^v_Y\left(y_{A,B}^v;t\right)$
\begin{align}
&\partial_t {\cal P}^v_Y\left(y_{A,B}^v;t\right)+\nabla \cdot {\bm {J^{*}}\left(y_{A,B}^v;t\right)}= - 2\omega_{\rm off}{\cal P}^v_Y\,,\label{e7}
\end{align}
with $J^*_\alpha=-m_y( k_B T\partial_\alpha {\cal P}^v_Y+{\cal P}^v_Y  \partial_\alpha W^{v*})$ without the summation convention,  and $\alpha = y^v_{A,B}$. Here
\begin{align}
W^{v*}(y_B^v-y_A^v)&\!=\!\!\!\int \!\!\diff x_A \diff x_B \frac{\chi\left(x_{A,B}, y_{A,B}^v\right) \, }{Z_Y\left(y_{A,B}^v\right)}\notag\\&\!\!\!\!\!\!\!\!\!\!\!\!\!\!\!\!\!\!\!\!\!\!\!\!\times W^v\left(x_{A,B}, y_{A,B}^v\right)\exp({- W^v/k_BT})\,\label{e8}
\end{align}
is the effective interaction between the two substrate regions $S_A$ and $S_B$. Note that $W^{v*}$ is a function of $(y_B^v-y_A^v)$ because the original interaction potential $W$ is a function of relative positions, such that the substrate is translational invariant. Due to this symmetry, it is instructive to perform a variable substitution:  $(y_A^v, y_B^v)\to(y_+,y_-)$, where $y_+ = y_A^v + y_B^v$ and $ y_- = y_B^v-y_A^v$. It is straightforward to verify that $y^+$ follows diffusional dynamics with mobility $m_y$, while the survival probability of $ y_-$ satisfies the following 1-variable FPE, 
\begin{align}
&\partial_t {\cal P}_{-}( y_-;t)+\partial_{y_-} {\bm {J_-}( y_-;t)}= - 2\omega_{\rm off}{\cal P}_-\,,\label{e19}
\end{align}
where $J_- = -2m_yk_B T\partial_{ y_-} {\cal P}_-+{\cal P}_{-}\partial_{y_-} W^{v*} $. Equation~(\ref{e19}) suggests that the distance between the two substrate regions follows the same equation of motion as a particle diffusing in 1D under the influence of the potential $W^{v*}$. The effective interaction $W^{v*}$ is essentially the averaged $W$ in the equilibrium distribution of $x_A$ and $x_B$. As shown in Fig.~\ref{fig4}(a), $W^{v*}$ becomes asymmetric for any rescaled substrate spring constant, $\tilde K= K d^2/\Delta E$, and its shape becomes more asymmetric for increasing $\tilde K$. When $\tilde K\to\infty$, the substrate reduces to a pure viscous substrate and the profile of $W^{v*}$ approaches that of $W^*$, the effective interaction for the viscous substrate (see discussion later in this section). In the $\tilde K\to 0$ limit in which the substrate is extremely soft, both $U_b^v$ and $W^{v*}$ become flat and there is no interaction between the two substrate regions,  resulting in vanishing contractility. However, for any finite value of $\tilde K$, the profile of $W{^*}$ will always be asymmetric for contraction and expansion, such that a positive contractile velocity is expected. Interestingly, a particle moving in a periodic $W^{v*}$ is mathematically equivalent to a motor binding on a polar filament~\cite{Julicher1997}, 
with the crucial distinction that the motor directional motion is dictated by the filament polarity, while the motion here is always contractile in character.

After obtaining the effective potential, we use Eq.~(\ref{e19}) to calculate the survival probability for $y_-$ and $\Delta y$ of Eq.~(\ref{e1}) is
\begin{equation}
\begin{aligned}
\Delta y  =  \langle y_-\rangle _{t=0}-\langle y_-\rangle _{t=t_e}\,,
\end{aligned}
\label{e20}
\end{equation}
where the average at $t=0$ and $t=t_e$ follows the same definition as in  Eq.~(\ref{e39}), with the average being respect to ${\cal P}_-$ instead of ${\cal P}^{\rm ve}$.

In Fig.~\ref{fig_ve}(c) we numerically calculate the contractile velocity as function of the rescaled substrate spring constant $\tilde K $. The velocity increases monotonically with $\tilde K$, as a result of the more asymmetric $W^{v*}$ for increasing $\tilde K$. For $K\gg \Delta E/d^2$, the velocity asymptotically converges to the velocity profile for a viscous substrate. Therefore, to reach a maximum contractile velocity, the substrate rigidity should be much larger than $ \Delta E/d^2$. 

The Maxwell substrate also introduces an intrinsic substrate relaxation time, $\tau_{s}=1/(Km_y)$. If $\tau_s\ll\tau_{\rm off}$, where $\tau_{\rm off}=1/(2\omega_{\rm off})$ is the average lifetime of the bound state, the substrate relaxes completely before the polymer unbinds, leading to a small contractile velocity. A monotonic dependence between the contractile velocity and the ratio $\tau_s/\tau_{\rm off}$ is observed in Fig.~\ref{fig_ve}(c). Clearly,  $\tau_s/\tau_{\rm off}$ must be large enough to reach a considerable contraction.  If $\tau_s/\tau_{\rm off}\gg 1$, the substrate behaves almost like an elastic substrate with spring constant $K$, with $ v/m_y$ being the contractile force exerted on the substrate. In this case the substrate relaxation can be neglected during a single binding/unbinding event, so the contractile force is not reduced by the substrate relaxation, leading to maximized contractile force. 

In Fig.~\ref{fig4}(b) we plot the relation between the contractile velocity and the binding site spacing $d$. Interestingly, we find that in the small-$d$ limit the velocities calculated from various $\tilde K$ values collapse on a single curve, which shows a quadratic dependence on $d$. This is not surprising because the microscopic model reduces to the coarse-grained model in this limit. In fact, when we analyze Eq.~(\ref{e8}) we find that the total energy $W^v$ is composed of two interactions, $U_e$ and $U_b^v$. The fluctuation lengthscale of $U_e$ is $\delta \ell$, while the distance between $x_A$ ($x_B$) and $y_A^v$ ($y_B^v$) is smaller than $d/2$, implying that  in the $d\ll \delta \ell$ limit we can approximate $U_e(x_B-x_A)$ with $U_e(y_B^v-y_A^v)$.  Furthermore, when $\Delta E^v\gg k_B T$ we have
\begin{equation}
\begin{aligned}
\exp(-U_b^v(x)/k_BT)\approx \left\{ 
\begin{aligned}
1 \qquad(x = nd)\\
0 \qquad(x \neq nd)
\end{aligned}
\right.
\end{aligned} \, ,
\label{e22}
\end{equation}
where $n$ is any integer. Substituting Eq.~(\ref{e22}) and $U_e(x_B-x_A)=U_e(y_B^v-y_A^v)$ into Eq.~(\ref{e8}) gives $W^{v*}=U_e$. This suggests that the 1-variable FPE  in Eq.~(\ref{e19}) is identical to Eq.~(\ref{e24}) in the coarse-grained model, with $y^-$ and $2m_y$ in Eq.~(\ref{e19}) replaced by $\ell_f$, $1/\gamma$ in Eq.~(\ref{e24}). Therefore, the microscopic model can be reduced to the coarse-grained model in the small-$d$ limit (when $\Delta E^v\gg k_B T$). In this case, the details of the effective binding potential $U_b^v$ are not important, and the contractile velocity of the microscopic model for any $K$ value is the same as that of the coarse-grained model. 

We also find that the contractile velocity shows a non-monotonic dependence on $d$ in Fig.~\ref{fig4}(b). The velocity increases with $d$ for $d\ll \delta \ell$,  decreases with $d$ for $d\gg \delta \ell$, and reaches its maximum at $d\sim \delta \ell$. When $d$ is large, the effective binding potential $U_b^v$ essentially becomes flat, weakening its ability to stretch or compress the polymer  when  it binds. This also explains the different predictions of the microscopic model and the coarse-grained model in the large-$d$ limit. The coarse-grained model is valid only when the change in $U_e$ is smaller than the binding potential, see discussion after Eq.~(\ref{e23}). Since the change in the polymer length is of the order of $d$, the change in $U_e$ can be approximated by $U_e(\ell_0+d)$, which increases dramatically with $d$ when $d>\delta \ell$. Therefore, for $d\gg \delta \ell$ we have $U_e \geq\Delta E^v$, which suggests that the effective binding potential  does not have  sufficient energy to deform the polymer, and the maximum velocity is always reached at $d\sim \delta \ell$. We also find that the value of $d$ at the maximum velocity is larger for increasing $\tilde K$ (Fig.~\ref{fig4}(b)), due to the fact that $\Delta E^v$ increases with $\tilde K$ (Fig.~\ref{fig_ve}(b)). 

Fig.~\ref{fig4}(b) implies that the values of $d$ and $\delta \ell$ needs to be close enough for considerable contraction. Biologically, the value of $d$ is fixed for a given substrate filament, which is about the same order of magnitude  as monomer size, {\it e.g.,} $d\approx 10 \rm nm$ for actin filaments. $\delta \ell$, on the other hand, has a strong dependence on the segment rest length $\ell_0$ (crosslinking distance),  $\delta \ell\sim\ell_0^2/\ell_p$. Therefore, we expect the contractile velocity of a crosslinked network to be maximized under an appropriate crosslinking distance. For actin cytoskeleton, for instance, a typical value of the crosslinking distance is $\ell_0\approx 1 \rm \mu m$, corresponding to $\delta \ell = 6.2 \rm nm$ and $\tilde d=1.6$. Surprisingly, we find this $\tilde d$ value is very close to the maximum-velocity position, see dashed yellow line in Fig.~\ref{fig4}(b). Such coincidence implies that the crosslinking distance in cytoskeleton may be the result of evolution that optimizes contractility.

\subsection{\it The limit of a viscous substrate}
\label{s3a}

Having discussed the microscopic model with a Maxwell substrate, let us switch to a simpler case in which the substrate is viscous, {\it i.e.}, the $K\gg\Delta E/d^2$ limit. 

Visually, a viscous substrate is obtained by removing the two springs in Fig.~\ref{fig_ve}(a). 
 In this case, the lengths of the springs $y_A^e$ and $y_B^e$ do not appear as variables of the system state, and the total potential energy is $W^v$ defined in Eq.~(\ref{e12}), with $U_b^v$ replaced by $U_b$. The survival probability, ${\cal P}^v\left(x_{A,B},y_{A,B}^v\right)$, is governed by the same FPE of Eq.~(\ref{e11}), with the initial condition Eq.~(\ref{e4}). 

Following the same steps of variable elimination in Eqs.~(\ref{e10}-\ref{e12}), the 4-variable FPE is reduced to a 1-variable equation governed by the potential energy $W^{*}(y_B-y_A)$, which has the same form as $W^{v*}(y_B^v-y_A^v)$ in Eq.~(\ref{e8}), with $U_b^v$ replaced by $U_b$.  As shown in Fig.~\ref{fig4}(c), $W^{*}$ becomes asymmetric for finite $d$, and its shape becomes more asymmetric for small $d$. In fact, in the small $d$ limit ($d\ll \delta \ell$), $W^{*}$ has the same form as the polymer PMF $U_e$, because in this limit the microscopic model reduces to the coarse-grained model (see discussion below Eq.~(\ref{e22})).

In Fig.~\ref{fig4}(d) we numerically calculate the contractile velocity and show its dependence  on $d$. It shows similar non-monotonically $d$-dependence as the viscoelastic substrate. This is because the viscoelastic substrate is equivalent to the viscous substrate with a modified binding potential, as we point out in Eq.~(\ref{e12}). We also find that the contractile velocity decreases for increasing $\tilde m_y$ (or decreasing $\omega_{\rm off}$), similar to that in the coarse-grained model (see Fig.~\ref{fig_min_v} (c)).

\section{Discussion}
\label{s5}
We have presented a detailed ({\it microscopic}) model that demonstrates how contractility can be produced in the absence of molecular motors. Our model considers the details of the binding/unbinding process of one polymer on a viscoelastic network, which is assumed to be a Maxwell material. We find that contractility naturally results from the active (un)binding that violates DB, together with the asymmetric force-extension relation that breaks spatial symmetry. Notably, both of these features are generic for biopolymer networks such as the cytoskeleton. In our model, the key lengthscale determining the contractile force, is the binding site spacing $d$. Taking $d\simeq10$nm, of order the size of a globular protein or the spacing of binding sites on a cytoskeletal filament, and $\delta\ell\simeq6$nm, corresponding to an actin filament of length $1\mu$m, our model predicts a maximum contractile force $\sim0.5 \rm pN$. As a comparison, single myosin molecule produces a force $\sim3 \rm pN$~\cite{Finer1994}, suggesting that our mechanism generates a weaker but comparable force relative to motors.  It is also worth emphasizing that our mechanism is additive in that it can generate such contractility among multiple filaments even in disordered networks. 

We have explored various limits of our detailed model. Of special importance are the limits in which the details of the binding process are neglected (the so-called {\it coarse-grained model}) which greatly simplifies the calculations and allow for intuitive understanding of the contractile mechanism. Such simplification is valid at $d\leq \delta \ell$, when  our coarse-grained model and microscopic model predict similar contractile velocities. However, their predictions differ when $d\gg\delta \ell$: For the microscopic model there exists an optimal $d$ value and the velocity decreases for $d$ exceeding that optimal value, while for the coarse-grained model the velocity always increases with $d$. This difference is due to the large elastic energy when stretching the polymer in the large-$d$-limit. In the coarse-grained model, the binding process deforms the polymer length with a binding probability, regardless of the energy required for the deformation. In the microscopic model, on the other hand, the deformation of the polymer length is limited by the binding potential, and the change in the elastic energy cannot exceed $\Delta E$. Therefore, in the large-$d$-limit the coarse-grained model fails. Similar to molecular motors, a large enough $\Delta E$ is required to generate considerable directional motion~\cite{Julicher1997}. 

In our microscopic model, we show that the binding/unbinding on a Maxwell substrate, is equivalent to the binding/unbinding on a viscous substrate with a modified binding potential. The reduction naturally originates from the definition of the Maxwell material, which is composed of a pure elastic  part and a pure viscous part. Since the elastic part responds to external stress immediately, it can always be regarded as a fast variable, allowing us to only consider the relaxation of the viscous part. Therefore, this result  does not rely on any specific assumption, nor is it limited to the particular model considered in this paper. Rather, it can be applied to any binding/unbinding events on Maxwell substrates, whether the binding/unbinding is in or out of equilibrium. Moreover, the binding site spacing of the modified binding potential remains unchanged for any substrate elastic rigidity. Because the binding site spacing is the only parameter associated with the substrate binding potential in the coarse-grained limit,  when $d\leq \delta \ell$ the microscopic model should be reduced to the  coarse-grained model with a viscous substrate, regardless of the original substrate elastic rigidity. 

In this work, we have focused on the Maxwell substrate, which behaves as solid at short times and fluid at long times.  Another well-known viscoelastic model is the Kelvin-vorgit substrate, which can be applied to our model as well. The Kelvin-vorgit substrate behaves as fluid in short time and solid in long time. Therefore, at short times we expect the Kelvin-Vorgit substrate to deform in the same way as the viscous substrate that has been discussed in Sec.~\ref{s3}, while at long times the substrate stops deformation due to its own elasticity. We choose not to study in detail the  Kelvin-Vorgit substrate because most biomaterials, including the cytoskeleton and the extracellular matrix, are known to fludize at long times~\cite{Broedersz2010, Ijima2018}. Therefore, the Maxwell model is more appropriate for the substrate we have in mind that are formed by  biomaterials. However, the Maxwell model does impose an instataneous elastic response of the substrate, while in general a viscoelastic material, such as the Burgers model, needs a finite time to build up its elastic stress. In our model we assume the elastic relaxation time of the substrate to be much smaller than the unbinding time such that we can consider it to be instataneous. 

We have only discussed the Maxwell substrate with a single relaxation time for simplicity, however, our mechanism should also generate contraction in Maxwell substrate with multiple relaxation times~\cite{wiechert1889,Broedersz2010, Chen2021}. Our discussion about the relaxation time (see Sec.~\ref{s3}) suggests that contractility will be observed as long as there exists one relaxation time that is larger than or comparable  to the unbinding time. 

The contractile mechanism proposed in our model requires activity that 
breaks  time-reversal symmetry together with a spatial asymmetry that directs the active motion. Such Brownian-ratchet-like mechanism is similar to the enzymatic cycle of molecular motors, which is fueled by the hydrolysis of ATP~\cite{Julicher1997}. However, there exists a distinct difference between the motor and non-motor mechanisms, which is the origin of the broken spatial symmetry. The asymmetry for motors comes from the geometrical polarity of the substrate filaments they bind to, which directs the motor motion from well defined plus (minus) end to the other end~\cite{Magnasco1993}. Because the motor motion is not directly related to contractility, it may require a specific network architecture to generate contraction~\cite{Ennomani2016}. On the contrary, the non-motor mechanism in this paper does not rely on substrate polarity. Instead, it relies on the mechanical asymmetry that is generic for any semiflexible biopolymer.  In addition to semiflexible polymers, polymers with small persistence length may also possess similar mechanical asymmetry in the opposite limit of semiflexibility, {\it e.g.,} DNA~\cite{Marko1994}.  Such asymmetry leads to a directed motion that is always towards the \emph{contractile direction}, resulting in contractility regardless of network architecture. Moreover, the non-motor mechanism may even generate contractility on apolar filaments, such as {\it intermediate filaments}, which has not been thought possible. 

Interestingly, although in this paper we have only considered the mechanical asymmetry on substrate filaments, the crosslinking proteins may also be responsible for such asymmetry if they are soft enough, {\it e.g.,} filamin. Contractility has been observed in disordered/apolar actin bundles when  the network is crosslinked with filamin~\cite{Weirich2017}, although the active, non-equilibrium aspect of this system is unclear. It is even possible that other apolar filaments like septin may be responsible for force generation, since they play a role in the contractile ring~\cite{Mavrakis2014,Valadares2017}. 

Despite the different origins of spatial asymmetry, the motor and non-motor mechanisms are not mutually exclusive. Rather, the non-motor mechanism should be considered as an additional mechanism to the motor mechanism. The non-motor mechanism is able to generate contractility on both polar and apolar substrates. In our model, we have assumed apolar substrates as a worst-case scenario in which motors do not work. However, considering a polar substrate within our model is straightforward, where one should replace the symmetric binding potential in  Sec.~\ref{s3} with an asymmetric one, in which case both the motor and non-motor mechanisms can take place simultaneously. In such a case, the non-motor mechanism enhances the contractility of the motor mechanism. The effects of the two mechanisms may be distinguished from their different $\tilde d$ dependence. The motor-driven contractility usually requires the polymer to buckle~\cite{Silva2011,Lenz2012}, which only happens when the tension exceeds $\tau_0$. Because $\tau_0\sim \delta \ell^{-1}$, the motor-driven contractility strengthens with decreasing $\tilde d=d/\delta \ell$. On the other hand, the contraction of non-motor mechanism, as we have shown for both coarse-grained and microscopic models, increases with increasing $\tilde d$ when $d\leq\delta \ell$, which shows opposite $\tilde d$-dependence to motor mechanism.

To conclude, our model may provide an explanation for recent observations of non-myosin-dependent dynamics of the contractile ring during cell division or cellularization \cite{Wloka2013, Xue2016}.
It is possible that ATP-dependent crosslinking by myosin in disordered actin networks or other structures may generate contractility~\cite{Ramaswamy2004,Shen2006,Wang2011,Silva2011,Alvarado2013,Tjhung2013,Markovich2019}, 
without the need for a motor \emph{power stroke}. 
Our model suggests a generic, steady-state mechanism for contraction even in apolar or fully disordered structures. 

\begin{acknowledgements}
	\emph{Acknowledgments}:
	This work was supported in part by the National
	Science Foundation Division of Materials Research
	(Grant No. DMR-1826623) and the National Science
	Foundation Center for Theoretical Biological Physics
	(Grant No. PHY-2019745). The authors acknowledge helpful discussions with A. Abdalla, M. Gardel, C. Schmidt and J. Theriot. 
\end{acknowledgements}

\appendix
\section{Estimation of $\tau_r$ and $\tau_{\rm hop}$}
\renewcommand{\theequation}{A\arabic{equation}}
\setcounter{equation}{0}
\renewcommand{\thefigure}{A\arabic{figure}}
\setcounter{figure}{0}
In this appendix we derive the expressions of $\tau_r$ and $\tau_{\rm hop}$ that are used in Sec.~\ref{s3}. $\tau_r$ and $\tau_{\rm hop}$ are two characteristic timescales of the modified binding potential $U_b^v$, whose height is $\Delta E^v$.  Since $U_b^v$ does not have an analytical expression, to obtain an estimation of the two timescales we approximate $U_b^v$ by a triangular potential with height $\Delta E^v$ and periodicity $d$. 
$\tau_r$ is the time required for the polymer end to relax within one binding site.  Let the binding site be within $(-d/2,d/2)$ and consider the diffusion of a particle with mobility $m_x$ within a triangular binding potential. We use the ``intrawell relaxation time'' introduced in Ref.~\cite{Derenyi1999} to estimate $\tau_r$. It is defined as the average mean-first-passage time of the particle from any fixed initial position $x_0$, to a final position $x$ that is sampled from a Boltzmann distribution governed by $U_b^v$:
\begin{align}
\tau_r &= \frac{1}{k_B T Zm_x}\int_{-d/2}^{d/2}\diff x\int_{x}^{d/2}\diff y\int_{y}^{d/2}\diff z\notag\\&\times\exp\Big[\left[-U_b^v(x)+U_b^v(y)-U_b^v(z)\right]/k_B T\Big]\notag\\&\approx \frac{ d^2}{2k_B Tm_x}\left(\frac{k_B T}{\Delta E^v}\right)\,,\label{A1}
\end{align}
where $Z = \int_{-d/2}^{d/2}\diff x \exp[-U_b^v(x)]$ is the partition function. In Eq.~(\ref{A1}), the integrals over $\diff y$ and $\diff z$ calculate the mean-first-passage time from $x_0$ to $x$, and the integral over $\diff x$ calculates the average mean-first-passage time with the Boltzmann weight of $U_b^v(x)$. Note that $\tau_r $ is independent of the initial position $x_0$.

$\tau_{\rm hop}$ is the average time for the polymer end to hop to another binding site. The time required to escape from a potential well can be estimated by the mean-first passage time from the bottom of the well to the top of the well, which is~\cite{Gardiner}:
\begin{align}
\tau_{\rm hop}&=\frac{1}{k_B Tm_x} \int_0^{d/2}\!\!\!\diff y\int_{-d/2}^y\!\!\!\diff z	\exp\Big[[U_b^v(y)-U_b^v(z)]/k_B T\Big]\notag\\&\approx \frac{d^2}{2k_B Tm_x}\exp\left(\frac{\Delta E^v}{k_B T}\right)\,.\label{A2}
\end{align}
We then conclude that for  $\Delta E^v\gg k_B T$, we have $\tau_r\ll\tau_{\rm off}\ll\tau_{\rm hop}$. It can be understood intuitively: in the large $\Delta E^v$ limit, the potential well is steep enough thus driving a fast relaxation within the potential well, while the high potential barrier prevents hopping towards another binding site.

\bibliography{citation}
\end{document}